# Ionic conductivity and relaxation dynamics in plastic-crystals with nearly globular molecules

D. Reuter, K. Seitz, P. Lunkenheimer[a], and A. Loidl

**AFFILIATION**

Experimental Physics V, Center for Electronic Correlations and Magnetism, University of Augsburg, 86135 Augsburg, Germany

[a] **Electronic mail:** peter.lunkenheimer@physik.uni-augsburg.de

**ABSTRACT**

We have performed a dielectric investigation of the ionic charge transport and the relaxation dynamics in plastic-crystalline 1-cyano-adamantane (CNA) and in two mixtures of CNA with the related plastic crystals adamantane or 2-adamantanon. Ionic charge carriers were provided by adding 1% of Li salt. The molecules of these compounds have nearly globular shape and, thus, the so-called revolving-door mechanism assumed to promote ionic charge transport via molecular reorientations in other PC electrolytes, should not be active here. Indeed, a comparison of the dc resistivity and the reorientational $\alpha$-relaxation times in the investigated PCs, reveals complete decoupling of both dynamics. Similar to other PCs, we find a significant mixing-induced enhancement of the ionic conductivity. Finally, these solid-state electrolytes reveal a second relaxation process, slower than the $\alpha$-relaxation, which is related to ionic hopping. Due to the mentioned decoupling, it can be unequivocally detected and is not superimposed by the reorientational contributions as found for most other ionic conductors.

## I. INTRODUCTION

The increasing use of renewable energy sources and the growing importance of electric vehicles is stimulating demands for more efficient, cheaper, and pollution-free energy-storage technologies. This has led to a tremendous boost of the research on battery technology and, indeed, considerable progress was achieved in recent years. Nevertheless, modern-day batteries (mostly of Li-ion type) and especially their three central components – anode, cathode, and electrolyte – still need significant improvements to fulfill all technical requirements.[1,2] For example, the currently employed electrolytes, which are mostly liquid, have a number of shortcomings like limited electrochemical stability, flammability, and leakage.[3] Replacing liquid electrolytes with solid materials could solve many of these issues.

Potential candidates for new types of solid-state electrolytes include superionic crystals,[4,5,6] polymers,[7,8,9] and so-called plastic crystals (PCs).[10,11,12,13,14,15,16,17,18,19,20,21,22] The latter, which are the topic of the present work, are molecular crystalline materials, exhibiting dynamic orientational disorder of the molecules.[23,24,25] These molecules are located on the sites of a three-dimensional lattice but possess rotational degrees of freedom, at high temperatures reorienting on short time scales. This reorientational motion often exhibits glassy freezing under cooling: Below the orientational glass-transition temperature $T_g^o$, where the characteristic relaxation time $\tau$ is of the order of 100 s, a so-called glassy crystal is formed (sometimes also termed "orientational glass") with essentially immobile, orientationally disordered molecules.[25,26,27,28] To allow for reorientational motions at $T > T_g^o$, the intermolecular interactions must be relatively weak, which causes most PCs to be rather easily deformable (hence the name "plastic crystals"[23]). This plasticity enables adaption to mechanical stresses which can be beneficial for application.

The PC electrolytes can be classified into two sub-groups: (i) Ionic PCs,[11,12,15,16,17,19,20,22,29] composed of cations and anions, of which at least one is sufficiently complex to allow for orientational degrees of freedom. (ii) Molecular PCs,[13,14,18,30,31,32,33] consisting of neutral molecules and a relatively small amount of admixed salt to provide ionic charge carriers. Obviously, the dynamic rotational disorder in PC electrolytes generates a high-entropy medium, which should be favorable for the translational ion hopping.[10,12,16,21,34,35] However, a general theory to explain the connection between orientational disorder of the molecules and the translationally moving ions for PC electrolytes is so far not available and might largely depend on the specific PC material and the ion species. In literature, several mechanisms are discussed. Some works assume a direct coupling of the ionic translation dynamics to the rotational motion of the asymmetric ions or molecules via a "paddle-wheel" or "revolving door" mechanism.[10,12,18,32,33,35] Here the ionic conductivity is assumed to be enhanced by transient free volume generated within the lattice by the molecular reorientations. An alternative approach explains the high ionic mobility in PCs based on the peculiar properties of the plastic-crystalline lattice: The weak intermolecular interactions, leading to the high plasticity and orientational disorder in PCs, make them prone to the diffusion of lattice defects and vacancies and/or plane slips.[36] Hence, the translational motions of ions could be mainly promoted by diffusion processes along such defects.[11,15,16,29,37]

Interestingly, for several PC electrolytes, including both ionic and molecular PCs, it was recently found that the ionic conductivity and the stability range of the plastic phase can be considerably enhanced by admixing a related molecular species of different size.[18,20,22,30,31,32,33] This was, e.g., demonstrated for succinonitrile [SN: $C_2H_4(CN)_2$], the most prominent representative of PC electrolytes with neutral molecules:[13,14] It revealed a strong conductivity enhancement when admixing glutaronitrile [GN: $C_3H_6(CN)_2$],[18,30] reaching up to three decades for samples with up to 80 mol% GN and with small amounts of added Li ions.[18] Similar results were also found for other SN-based mixtures.[33] Mixtures of the PCs cyclohexanol and cyclooctanol also revealed a strong conductivity variation, depending on the mixing ratio.[32]

If the revolving-door mechanism indeed is dominating the ionic charge transport in such systems, one would expect a close coupling of the molecular rotational and ionic translation dynamics. This can be checked, e.g., by comparing the temperature dependences of the ionic dc conductivity and the reorientational relaxation times, which both can be determined by dielectric spectroscopy. For the SN-based systems, such close coupling was only revealed for certain mixtures, e.g., with very large amounts of a somewhat bigger molecule (like in $SN_{0.2}GN_{0.8}$) while for many other cases (e.g., $SN_{0.85}GN_{015}$) much weaker coupling was found.[18,33] In marked contrast, for the mentioned cyclohexanol-cyclooctanol mixtures, the coupling was very close for all mixing ratios.[32] Even the pure systems showed strong coupling while pure SN exhibits complete decoupling.[18] These findings are consistent with the revolving-door scenario: For the more-or-less disc-shaped cyclo-alcohol molecules, which reorient isotropically, during short time intervals widely opened gaps can occur, corresponding to parallel orientations of adjacent molecules. Through these gaps the ions can easily pass, implying an highly effective revolving-door mechanism which dominates the charge transport. In contrast, for the bulkier SN molecules the gaps opened during rotation obviously should be smaller and, thus, less favorable for the ionic charge transport. Here only the admixture of a considerable amount of the larger and less bulky GN molecules leads to effective revolving doors.[18]

To check this notion, in the present work we investigate the ionic charge transport and reorientational dynamics in 1-cyano-adamantane (CNA) and its mixtures with 20 mol% adamantane (ADA) or 2-adamantanon (AON). We introduced Li$^+$ ions into these mixtures by adding 1% Lithium bis(trifluoromethane)sulfonimide (LiTFSI), an often-used salt for electrolytes. The pure materials are well-known PCs and CNA and AON have a dipolar moment allowing for a dielectric investigation of their molecular dynamics.[25] The molecules of the three materials are rather close to a globular shape. Therefore, the gaps opened during their rotation should be almost negligible and the revolving-door mechanism should not play a significant role. Consequently, complete decoupling of the ionic diffusion from the rotational motions is expected for these PCs, which would confirm the revolving-door picture proposed for the less globular PC electrolytes. We want to remark that the present work does not aim at finding materials with high ionic conductivity. suitable for application, but instead intends to contribute to a better understanding of the applicability of the revolving-door mech for ionic conductorsanism.

The ADA molecule ($C_{10}H_{16}$) is a highly symmetrical non-dipolar carbon cage consisting of ten carbon atoms with the free bonds being saturated by hydrogen atoms. Below its melting temperature $T_m = 543$ K, ADA crystallizes in a PC phase with face-centered cubic (fcc) structure[38] and undergoes an order-disorder transition at 208 K.[39] With its highly symmetric, nearly globular molecules, ADA is a prototypical molecular PC. CNA ($C_{10}H_{15}CN$) consist of the same carbon cage as ADA with the addition of a cyano group. At $T_m \approx 460$ K the melt crystallizes in a cubic fcc PC phase.[40] At $T < 280$ K a transition into an orientationally ordered crystal occurs. The PC phase of CNA can be supercooled. Values for $T_g^o$ between about 163 and 178 K were reported,[25,41,42,43,44] (see Ref. 43 for a discussion of problems in the determination of $T_g^o$ for this PC). AON is of similar molecular structure, but the cyano group is replaced by an oxygen atom, connected by a double bond to the carbon cage. Below its melting point $T_m \approx 529$ K, AON forms a plastic-crystal with fcc structure.[45] It undergoes a partial ordering transition at about 180 K, below which orientational motions are still possible.[25,46] $T_g^o$ of this phase is about 131 K.[25]

## II. EXPERIMENTAL METHODS

CNA was purchased from AppliChem (purity 99.8%). The other chemicals ADA (purity > 99%), AON (purity 99%) and LiTFSI were obtained from Sigma-Aldrich. All chemicals were used as received. For the mixtures, suitable amounts of ADA, AON, and LiTFSI were dissolved in melted CNA at about 470 K. Differential scanning calorimetry (DSC) was performed with a DSC 8500 from Perkin Elmer with a scanning rate of 10 K/min.

For the dielectric measurements, liquid samples were filled into preheated parallel-plate capacitors with a plate distance of 0.1 mm. The measurements were carried out using a frequency-response analyzer (Novocontrol Alpha). For temperature control a $N_2$-gas cryostat (Novocontrol Quatro) was employed. All measurements were done under cooling. Prior to the measurements, the samples were dried in vacuum atmosphere for four hours. During drying, water removal was monitored with periodical dielectric sweeps. After an initial drop, the conductivity settled quickly on a constant value, marking the effective removal of water.

## III. RESULTS AND DISCUSSION

### A. DSC results

To check for the phase- and glass-transition temperatures of the mixtures, we performed DSC measurements. As revealed by Fig. 1(a), for $CNA_{0.99}LiTFSI_{0.01}$ the applied precooling with a rate of 10 K/min obviously was sufficient to supercool the mentioned low-temperature solid-solid transition of pure CNA and a glassy crystal was obtained. During the heating run, shown in Fig. 1(a), an orientational glass transition shows up at $T_g^o \approx 167$ K (deduced using the onset evaluation method), revealed by the typical sigmoidal increase in the DSC heat flow [inset of Fig. 1(a)]. The orientational glass transition is succeeded by a cold-crystallization into the ordered phase at about 220 K, which melts again at $T \approx 270$ K into the PC phase. The final melting into the liquid occurs around 460 K. For the mixtures with 20 mol% of ADA and AON, also clear glass transitions are detected with $T_g^o \approx 157$ K and 156 K, respectively. No indications of solid-solid transitions were found in these mixtures. These findings are in accord with other works on mixed PC system, where an extension of the PC phase down to lower temperatures with the addition of the second compound was found.[18,20,21,47]

temperatures, $T \geq 304$ K, a strong additional increase towards low frequencies occurs (not completely shown in the figure), leading to unrealistic values of $\varepsilon'$ exceeding $10^3$. It can be ascribed to non-intrinsic electrode-polarization effects arising from the accumulation of ionic charge carriers at the electrodes as often found for ionic conductors.[48]

The dielectric-loss spectra are dominated by a peak, shifting from lower to higher frequencies with increasing temperatures [Fig. 2(b)]. The peak positions match the points of inflection of the faster sigmoidal decrease in $\varepsilon'(\nu)$. Overall, these spectral features reveal the typical signatures of a relaxation process arising from molecular reorientations. When comparing the spectra with literature data,[25,43,49,50,51] indeed it can be identified as the main reorientation process, i.e., the $\alpha$ relaxation, of CNA. The loss-peak frequency $\nu_p$ is related to the average relaxation time $\langle \tau \rangle$ of the $\alpha$ process via $\langle \tau \rangle \approx 1/(2\pi \nu_p)$. Its continuous shift towards lower frequencies with decreasing temperature reflects the slowing down of molecular motion under cooling, typical for glassy freezing.

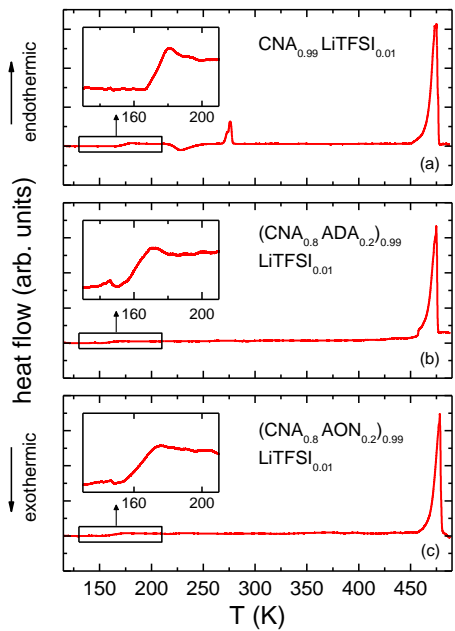

**FIG. 1.** DSC heat flow measured under heating from 110 K to 490 K with rates of 10 K/min for the three investigated mixtures. Prior to these measurements, the samples were cooled from 490 K to 110 K with 10 K/min. Endothermic processes are plotted in positive y-direction. The insets magnify the data in the temperature region of the glass transition.

## B. Dielectric spectra

Figure 2 shows the spectra of the dielectric constant $\varepsilon'$ (a), the dielectric loss $\varepsilon''$ (b), and the conductivity $\sigma'$ (c) for the pure CNA sample with 1% LiTFSI. The measurement was performed during cooling from 376 K with 0.4 K/min. The $\varepsilon'$ spectra reveal a double step-like decrease with increasing frequency. For example, for 304 K the slower sigmoidal decrease is located at about 10 - 100 Hz and the faster one at about $10^6$ Hz. With increasing temperature, both steps continuously shift to higher frequencies and their separation becomes somewhat reduced. For the highest

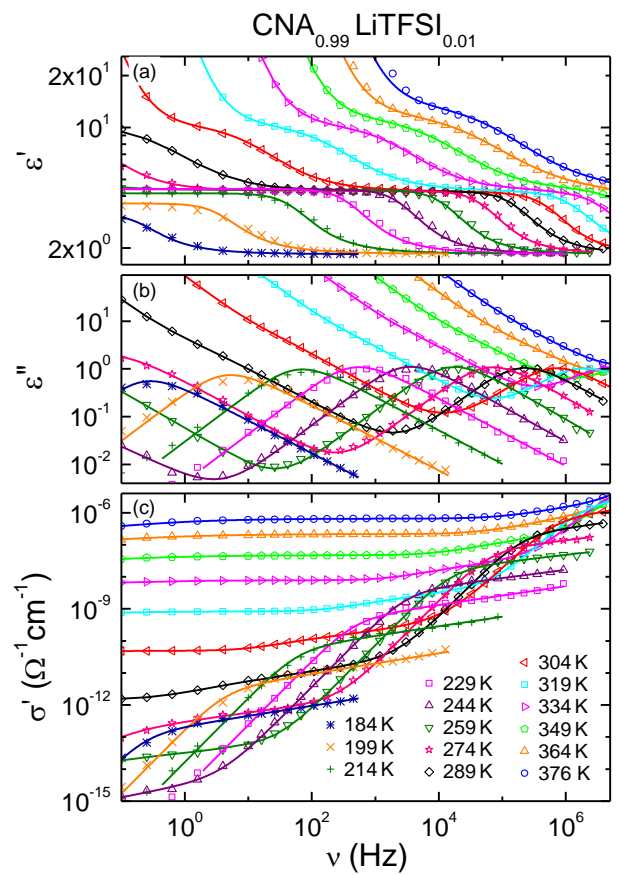

**FIG. 2.** Frequency-dependent dielectric constant (a), dielectric loss (b), and conductivity (c) of $CNA_{0.99}LiTFSI_{0.01}$ for various temperatures. The solid lines in (a) and (b) are fits, simultaneously performed for $\varepsilon'$ and $\varepsilon''$, using an equivalent-circuit approach,[48,52] consisting of a distributed RC circuit to describe the electrode-polarization effects,[48] a dc-conductivity contribution, and two relaxation functions (see text for details). The lines in (c) were calculated from the fits of $\varepsilon''$.

At frequencies $\nu \ll \nu_p$, a marked increase shows up in $\varepsilon''(\nu)$ with slope -1, implying $\varepsilon'' \propto 1/\nu$ in this region. It can be attributed to the ionic dc conductivity $\sigma_{dc}$ of the sample. Via the relation $\varepsilon'' = \sigma'/(2\pi\nu\varepsilon_0)$ (with $\varepsilon_0$ the permittivity of vacuum), this feature corresponds to a frequency-independent plateau in $\sigma'(\nu)$ as observed for low frequencies and higher temperatures in the conductivity spectra of Fig. 2(c). For example, at 334 K this dc plateau extends from the lowest frequencies up to about $10^3$ Hz. However, at the three highest temperatures shown in Fig. 2(c), at low frequencies the conductivity exhibits an additional weak decrease, which is due to the electrode-polarization effects mentioned above.[48]

Between the dc-conductivity contribution and the $\alpha$-relaxation, the slope -1 of the $\varepsilon''(\nu)$ curves [Fig. 2(b)] crosses over to a more shallow decrease. Correspondingly, with increasing frequency, following the $\sigma_{dc}$ plateau, the $\sigma'$ spectra in Fig. 2(b) reveal a slight increase before the onset of the $\alpha$-relaxation contribution, the latter appearing as a pronounced shoulder in this quantity. This additional spectral feature arises from the loss peak associated with the mentioned slower relaxation process, which is clearly evidenced by the low-frequency step in the $\varepsilon'$ spectra of Fig. 2(a). This peak, however, is strongly superimposed by the dc-conductivity contribution, becoming dominant in $\varepsilon''$ and $\sigma'$ at low frequencies. The physical origin of this process will be discussed below.

To unequivocally deduce $\sigma_{dc}(T)$ and $\tau(T)$ of both relaxation processes from the dielectric spectra, we have simultaneously fitted $\varepsilon'(\nu)$ and $\varepsilon''(\nu)$ applying an equivalent-circuit approach.[48,52] As previously demonstrated for various ionically conducting materials,[48,53,54,55] including PC electrolytes,[18,32,33] a distributed RC circuit connected in series to the bulk sample provides a good formal description of electrode polarization. To account for the $\alpha$ relaxation, we used the phenomenological Cole-Davidson (CD) function[56] as previously done for pure CNA.[25,51] For the additional slower relaxation, good fitting results were reached employing the symmetrically broadened Cole-Cole (CC) function.[57] The dc-conductivity contribution in the loss was accounted for by $\varepsilon''_{dc} = \sigma_{dc}/(2\pi\nu\varepsilon_0)$. A good agreement of fits and experimental data could be achieved in this way (lines in Fig. 2). One should be aware that, depending on temperature, only part of the elements of the overall fit function had to be used for the fits (e.g., at low temperatures the electrode polarization plays no role), thus avoiding an excessive number of fit parameters.

Figure 3 presents the same data as Fig. 2 for three selected temperatures only. To help clarifying the physical origin of the slower relaxation process found in $CNA_{0.99}LiTFSI_{0.01}$, a spectrum measured at 319 K for pure CNA without any added Li salt is included in this figure (crosses). In $\varepsilon'$ [Fig. 3(a)], the Li-free CNA result perfectly matches the corresponding $CNA_{0.99}LiTFSI_{0.01}$ data at high frequencies, in the region of the $\alpha$-relaxation. However, while Li-doped CNA exhibits the second relaxation below about $10^4$ Hz, the ion-free pure CNA does not display such an additional contribution. Instead, it only shows a frequency-independent $\varepsilon'$ in this region, corresponding to the static dielectric constant of the $\alpha$ relaxation. In accord with these findings, the contributions from the detected slower relaxation in the $\varepsilon''$ and $\sigma'$ spectra also are absent in the Li-free CNA sample [Figs. 3(b) and (c)]. For 319 K, in $CNA_{0.99}LiTFSI_{0.01}$ these contributions are detected between about $10^2$ and $3\times10^4$ Hz. In ion-free CNA, in this frequency region the $\varepsilon''$ spectrum instead only reveals the left flank of the $\alpha$ relaxation, before the dc-conductivity contribution sets in below about $5\times10^2$ Hz. Overall, this comparison with Li-free CNA indicates that the slow relaxation-like process in $CNA_{0.99}LiTFSI_{0.01}$ is related to the presence of the added ions. It should be noted that the dc conductivity in the Li-free sample probably arises from marginal amounts of ionic impurities. As expected, Fig. 3(c) reveals that it is several decades lower than for the Li-doped sample.

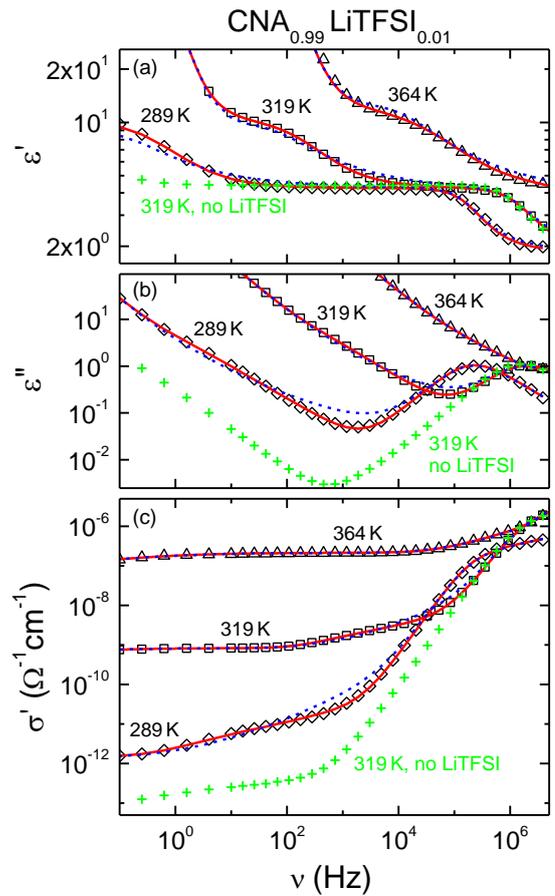

**FIG. 3.** Comparison of different fit functions to account for the slower relaxation in $CNA_{0.99}LiTFSI_{0.01}$. As typical examples, we show spectra of the dielectric constant (a), dielectric loss (b), and conductivity (c) at three selected temperatures. The solid lines are the same as in Fig. 2 assuming a CC function for this process. The dashed lines are fits using the RBM (newer version from Ref. 71) to describe the slow relaxation (see text for details). The crosses show spectra at 319 K for CNA without added Li salt, revealing the absence of the slow relaxation in the pure sample. These data were vertically scaled to match the $\varepsilon''$-peak amplitude of the corresponding spectrum for $CNA_{0.99}LiTFSI_{0.01}$ (same factor applied for all three quantities).

There are several possibilities to explain the ion-related slow relaxation process detected in $CNA_{0.99}LiTFSI_{0.01}$. For example, it might emerge from the reorientation of the dipolar TFSI anions of the added LiTFSI. However, as only 0.5 mol% of the sample are TFSI anions and as the amplitude [i.e., the step height in $\varepsilon'(\nu)$] of the slow relaxation is of similar order as for the $\alpha$ relaxation, this explanation seems unlikely. Another scenario could be ions, attached to the CNA molecules. The resulting increased steric hindrance and enhanced dipolar moment may explain the slower relaxation time and relatively high relaxation strength, respectively. Moreover, this relaxation could be of non-intrinsic Maxwell-Wagner type,[58,59] arising from the grain boundaries within the polycrystalline sample.[52] A similar space-charge contribution to the dielectric spectra was proposed in Ref. 30 for a SN-GN mixture, however, based on the assumption of partial phase separation into orientationally ordered and disordered regions.

Another interesting possibility is the relaxation-like process, sometimes assumed to universally occur in dielectric permittivity spectra due to ionic diffusion.[60,61,62,63,64] A prominent example is a small relaxation process found in $\varepsilon'$ measurements of the ionic-melt glass former $[Ca(NO_3)_2]_{0.4}[KNO_3]_{0.6}$ (CKN), which does not contain any dipolar molecules or ions.[60,65] Within the modulus fomalism for the description of ionic conductors, its occurrence is rationalized by a distribution of conductivity-relaxation times.[60] However, one should be aware that the usefulness of this formalism is controversially discussed.[66,67,68,69] An alternative explanation of such an ionic-diffusion related relaxation observed in the permittivity is provided by the random free-energy barrier hopping model (RBM), which accounts for ionic hopping conduction on a microscopic level.[70,71] This model predicts relaxation-like steps in $\varepsilon'(\nu)$ that reflect purely translational ionic dynamics and are not related to any reorientational motion. In Fig. 3, fits using the most current version of the RBM[71] to model the ionic relaxation (dashed lines) are compared to the fits using the empirical CC function also shown in Fig. 2 (solid lines). For these fits, in our equivalent-circuit approach the CC function and the dc conductivity were replaced by the prediction of the RBM. This approach at least qualitatively describes the detected slow relaxation. At the lowest presented temperature, some deviations in the region between the two relaxation processes show up, especially in $\varepsilon''$. Since the RBM effectively uses two parameters less than the CC plus dc-conductivity approach, such deviations could be expected. Overall, the RBM, an ionic hopping model, is able to roughly describe the slow relaxation process, supporting the notion that the slow relaxation is of ionic origin.

The comparison of the conductivity spectra of CNA without and with the addition of 1% Li salt [Fig. 3(c)] evidences a massive (~ four decades at 319 K) enhancement of the dc conductivity when a rather small amount of ions is added. Thus, as found for several other PCs, CNA can provide a solid matrix for ionic charge transport. Of course, the detected dc conductivity for the given 1% Li-salt concentration is too low for any application. Previous works from our group have established that the ionic conductivity in such ion-doped PC electrolytes can be increased by admixing a related compound.[18,32,33] In SN, the addition of the larger GN, adiponitrile, or pimelonitrile molecules revealed marked conductivity enhancements of up to three decades, whereas the enhancement effect increased with growing molecular size[33] and concentration.[18] Such an enhancement was also found for the admixture of malononitrile, which has smaller molecules than SN. All these findings can be qualitatively understood within the revolving-door scenario mentioned in the introduction. In the following, we check whether this effect is also found in mixed PC systems based on CNA.

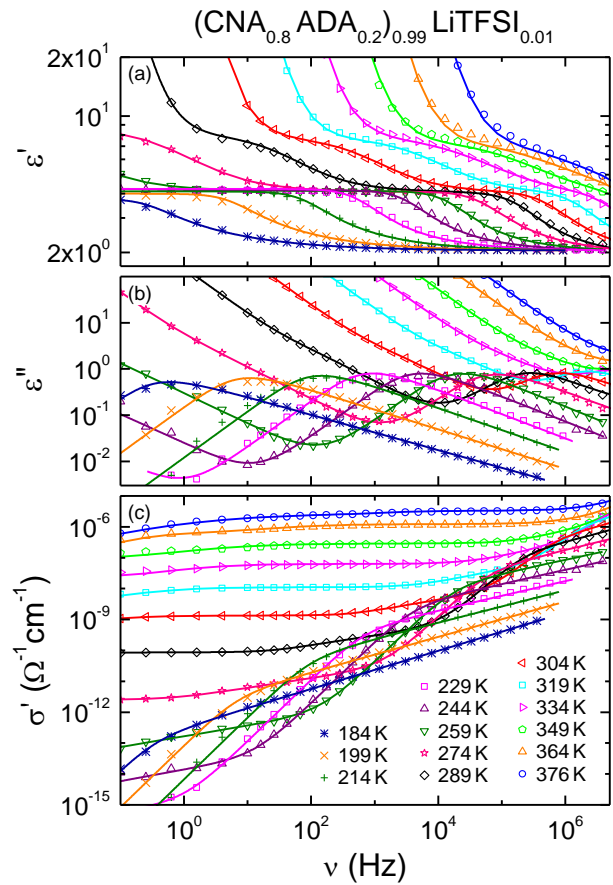

**FIG. 4.** Frequency-dependent dielectric constant (a), dielectric loss (b), and conductivity (c) of $(CNA_{0.8}ADA_{0.2})_{0.99}LiTFSI_{0.01}$ for various temperatures. The lines are fits with the same approach as in Fig. 2.

For this purpose, dielectric measurements of the mixtures $(CNA_{0.8}ADA_{0.2})_{0.99}LiTFSI_{0.01}$ (Fig. 4) and $(CNA_{0.8}AON_{0.2})_{0.99}LiTFSI_{0.01}$ (Supplementary Material) were performed. At first glance, the spectra in Fig. 4 closely resemble those for $CNA_{0.99}LiTFSI_{0.01}$ (Fig. 2). In $\varepsilon'$, a double-relaxation step shifts from low to high frequencies with increasing temperature, accompanied by a peak in $\varepsilon''$. Significant dc-conductivity contributions are evidenced by plateaus in $\sigma'$ and the corresponding $1/\nu$ power laws in $\varepsilon''$ at frequencies below the peaks (only partly shown). Weak deviations from these dc contributions evidence the superimposed loss peaks associated with the slower relaxation (e.g., around 100 Hz for 289 K). Moreover, electrode-polarization effects emerge at low frequencies for the highest temperatures. Consequently, the interpretation of the involved processes and the fitting routine are the same as for $CNA_{0.99}LiTFSI_{0.01}$. The corresponding fits, shown by the lines in Fig. 4, are in reasonable agreement with the experimental data and allow for the precise determination of the relaxation time and dc conductivity.

A closer look at Fig. 4 reveals some minor differences to the $CNA_{0.99}LiTFSI_{0.01}$ data of Fig. 2: For the highest temperature in the ADA mixture (376 K), the onset of electrode polarization in $\varepsilon'$ occurs at about $10^5$ Hz, which is roughly one decade higher than in $CNA_{0.99}LiTFSI_{0.01}$. This is due to the about one decade higher dc conductivity of the ADA mixture, which becomes directly obvious when comparing the dc plateaus in Figs. 2(c) and 4(c). Another noticeable difference is the extremely broadened loss peak of the mixture. For the lowest temperature, $T = 184$ K, the right flank of the peak can be observed over six frequency decades before falling below the measurement limit of about $10^{-2}$. Interestingly, the peaks at all temperatures can still be described by a single CD function. Qualitatively similar findings as for $(CNA_{0.8}ADA_{0.2})_{0.99}LiTFSI_{0.01}$ were also obtained for $(CNA_{0.8}AON_{0.2})_{0.99}LiTFSI_{0.01}$ (Supplementary Material, Fig. S1). An increased broadening of the $\alpha$-relaxation peaks in the mixtures, which obviously must have higher disorder due to the statistical distribution of the added ADA or AON molecules, seems reasonable when considering the common explanation of broadened relaxations features by a distribution of relaxation times.[72,73]

## C. DC conductivity, relaxation times, and their interrelation

For a direct comparison of the dc conductivities of the three investigated samples, Fig. 5 provides an Arrhenius plot of their temperature-dependent $\sigma_{dc}$ as obtained from the fits shown in Figs. 2, 4, and S1. The conductivities of the two PC mixtures are significantly higher than for $CNA_{0.99}LiTFSI_{0.01}$. Around room temperature, the admixture of 20 mol% ADA or AON leads to an enhancement of $\sigma_{dc}$ by about 1.5 or 2 decades, respectively. With decreasing temperature, the differences in the conductivity become more pronounced. Thus, just as previously reported for the SN-based mixtures and for the cyclohexanol-octanol system,[18,30,32,33] admixing a related compound to CNA also leads to a considerable enhancement of the ionic dc conductivity.

As revealed by Fig. 5, $\sigma_{dc}(T)$ of all samples exhibits clearly nonlinear behavior, evidencing significant deviations from simple thermally activated charge transport, $\sigma_{dc}(T) \propto \exp[-E/(k_BT)]$. Non-Arrhenius behavior is typical for glassy dynamics and found, e.g., for the structural relaxation of glassforming liquids, where it is usually well fitted by the empirical Vogel-Fulcher-Tammann (VFT) equation.[74,75,76,77] For the ionic dc conductivity in materials exhibiting glassy freezing,[53,54,78,79] including PCs,[18,32,33] such non-Arrhenius behavior also is often found. (One should note that a sufficiently broad temperature range is necessary to detect the deviations from Arrhenius.) It can be described by the conductivity variant of the VFT equation,

$$\sigma_{dc} = \sigma_0 \exp\left[\frac{-DT_{VF}}{T - T_{VF}}\right] \qquad (1).$$

Here $\sigma_0$ represents a pre-exponential factor, $D$ is the so-called strength parameter that quantifies the deviation from Arrhenius behavior,[77] and $T_{VF}$ is the Vogel-Fulcher temperature, where $\sigma_{dc}$ should become zero. The lines in Fig. 5 are fits using this equation; they reasonably describe the experimental data. We obtain $D = 14.9$ for the CNA system, 20.6 for the CNA-ADA mixture, and 12.5 for the CNA-AON mixture, leading to values of the so-called fragility index[80] of $m \approx 56$, 45, and 63, respectively. This corresponds to intermediate fragility within the strong/fragile classification scheme of glassy freezing.[77,80]

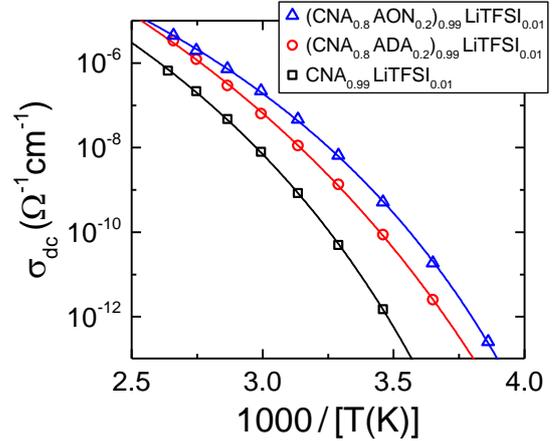

**FIG. 5.** Temperature-dependence of the dc conductivity of all three investigated mixtures (Arrhenius representation). The values of $\sigma_{dc}$ were obtained from the fits of the dielectric spectra shown in Figs. 2, 4, and S1. The lines are fits with Eq. (1).

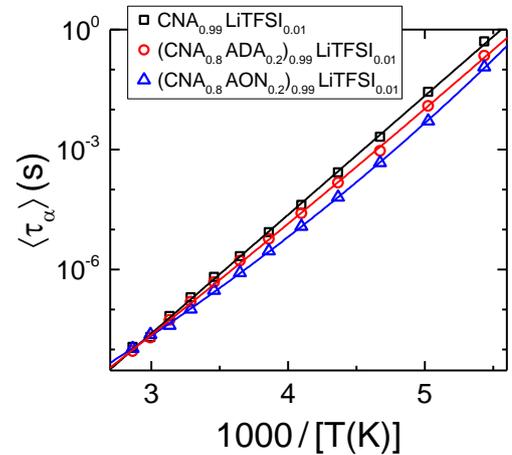

**FIG. 6.** Temperature dependence of the average dipolar relaxation times of the three investigated mixtures, plotted in Arrhenius representation. The lines are fits using an Arrhenius law for $CNA_{0.99}LiTFSI_{0.01}$ and VFT laws for the other two samples.

What is the microscopic reason for the conductivity enhancement documented in Fig. 5? In previous investigations of PC electrolytes, similar mixing-induced conductivity enhancements could either be rationalized by an increased coupling of molecular reorientational and ionic translational motion[18,33] or by the acceleration of the molecular relaxation dynamics under constant coupling.[32] The results of Fig. 6, comparing the reorientational $\alpha$- relaxation times $\tau_\alpha$ of the three investigated materials, reveals that the latter explanation is not applicable in the present case: At all temperatures, the relaxation times of the three materials differ by less than one decade. Especially, in the temperature range of Fig. 5 the difference in $\tau$ is significantly smaller than the 1.5 or 2 decades variation found for the dc conductivity. Figure 6 also reveals no (for $CNA_{0.99}LiTFSI_{0.0}$) or only weak deviations of $\tau_\alpha(T)$ from Arrhenius behavior (for the other two systems; $m \approx 18$ for CNA-ADA and 25 for CNA-AON) as often found for PCs.[25,47,78] In contrast, for $\sigma_{dc}(T)$ pronounced deviations from Arrhenius are detected for all samples (Fig. 5).

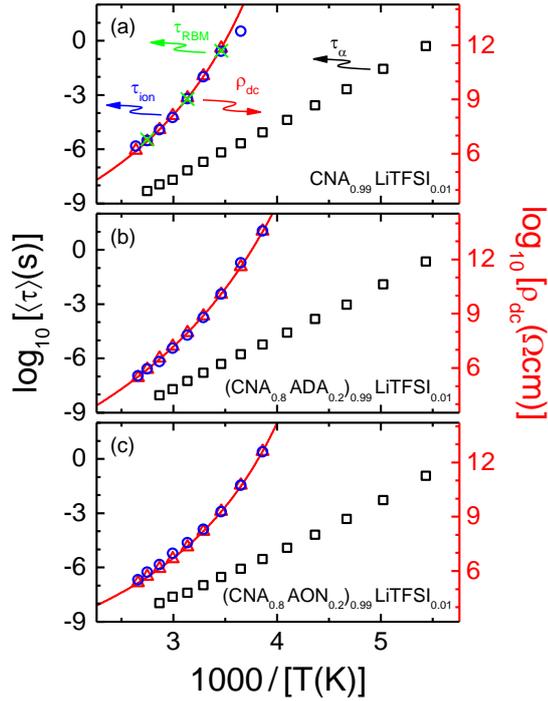

**FIG. 7.** Temperature dependences of the average relaxation times and of the dc resistivity of $CNA_{0.99}LiTFSI_{0.01}$ (a), $(CNA_{0.8}ADA_{0.2})_{0.99}LiTFSI_{0.01}$ (b), and $(CNA_{0.8}AON_{0.2})_{0.99}LiTFSI_{0.01}$ (c). The reorientational $\alpha$-relaxation time (squares), slow relaxation time (circles), RBM relaxation time (crosses) and dc resistivity (triangles) were obtained from the fits described in the text. The y-axis ranges of $\langle\tau\rangle$ (left ordinates) and $\rho_{dc}$ (right ordinates) were adjusted to cover the same number of decades. The starting values of the ordinates were chosen to achieve a match of the relaxation time of the slow relaxation and the dc resistivity. The solid lines through the $\rho_{dc}(T)$ data are calculated from the VFT fits of $\sigma_{dc}(T)$ presented in Fig. 5.

Figure 7 enables a direct comparison of the ionic translational and molecular reorientational dynamics in the present samples. For the three investigated systems, it shows the temperature-dependent average relaxation times of the two relaxation processes (squares and circles, left ordinates) and the dc resistivities $\rho_{dc} = 1/\sigma_{dc}$ (triangles, right ordinates) within the same Arrhenius plot. By ensuring an identical number of decades on both ordinates in Fig. 7, vertical shifts allow for a direct comparison of the relative temperature variation of both quantities. In contrast to our findings in several other (but not all) PC electrolytes,[18,32,33] the reorientational $\alpha$-relaxation times in these CNA-based systems cannot be scaled on the dc resistivity. As already revealed by Figs. 5 and 6, the deviations from Arrhenius temperature dependence are nearly negligible for the $\alpha$-relaxation time while they are considerable for the resistivity (and, thus, for the conductivity). In addition, the overall temperature dependence of $\rho_{dc}$ is much stronger than that of the $\alpha$-relaxation time for all samples. Obviously, in these systems the ionic charge transport is completely decoupled from the molecular reorientations and the revolving-door mechanism, considered for other conducting PCs, does not play any role. The coupling does not change when 20 mol% of a related PC compound are added to CNA and a variation of coupling cannot explain the differences in $\sigma_{dc}$ documented in Fig. 5. It is an interesting finding that the decoupled ionic charge transport in these PCs reveals such pronounced non-Arrhenius behavior. This may point to a glasslike behavior of the ionic subsystem as previously considered for other ionic conductors.[78]

The circles in Fig. 7 show the relaxation times $\tau_{ion}$ of the detected additional slow relaxation process (Figs. 2, 4, and S1), which is absent in ion-free CNA (Fig. 3). In marked contrast to the $\alpha$ relaxation, they can be almost perfectly scaled to the dc-resistivity curves (triangles) for all three materials. This confirms that this relaxation process indeed is due to ionic dynamics as discussed in detail above. The relaxation times $\tau_{RBM}$, obtained from the fits with the RBM shown in Fig. 3 for $CNA_{0.99}LiTFSI_{0.01}$, also well scale with $\rho_{dc}(T)$ [Fig. 7(a), crosses], further supporting the ionic nature of the slow relaxation.

The reader may note that the overall appearance of the relaxation-time maps in Fig. 7 reminds of the characteristics of primary $\alpha$ and secondary $\beta$ relaxational processes, often found for glass-forming liquids.[81,82,83] There the slower primary relaxation shows super-Arrhenius behavior and is related to the structural relaxation, while the faster secondary relaxation, termed Johari-Goldstein relaxation,[81] closely follows Arrhenius behavior at low temperatures. However, it seems difficult to imagine how this picture could account for the present findings: Here the two relaxation processes arise from different constituents of the samples (hopping ions and rotating molecules) and the slower relaxation, which would be the $\alpha$ relaxation within this framework, only appears with the addition of 1% ions (cf. Fig. 3).

## IV. SUMMARY AND CONCLUSIONS

In summary, our results clearly point to a complete decoupling of the ionic charge transport in the investigated CNA-related PCs from the reorientational molecular motions present in these systems. Obviously, the revolving-door mechanism, discussed for other plastic-crystalline solid-state electrolytes,[10,12,18,32,33,35] does not play any role here. Interestingly, this finding in principle supports the idea that this mechanism is active in the other PC systems, like the SN-based mixtures mentioned in the introduction. These PCs, with much higher conductivity than the present materials, are formed by more asymmetric molecules, whose rotation opens large gaps for passing ions. In contrast, the more globular molecules of the present systems do not lead to significant gaps during reorientation. Consequently, the ionic motions in these PCs are completely independent from the molecule reorientations.

Similar to the SN-related[18,30,33] and the mentioned cyclohexanol-octanol PC mixtures,[32] CNA also reveals a considerable increase of the dc conductivity when admixing a related PC compound (Fig. 5). However, in this case a revolving-door related mechanism as invoked for explaining this effect in the other PCs,[18,33,32] obviously cannot account for this finding. Most likely the ionic charge transport in these PCs is related to defects occurring in the plastic-crystalline lattice as previously considered for conducting PCs.[14,18,29,37,84] The addition of 20 mol% of a different molecule, in principle introduces a large number of additional defects, thus increasing the conductivity. Such defect-related ion mobility, of course, also should be present in the other PCs but, in cases where a good coupling to the reorientations is found,[18,32,33] this presumably is not the main charge-transport mechanism. The significantly lower room-temperature conductivity of the present CNA-related systems, compared to the nitriles[18,33] and the cyclohexanol-octanol PCs,[32] both with comparable ion concentrations, is compatible with this scenario.

Finally, we want to remark that the investigated CNA-based PC electrolytes represent nice examples where the relaxational dynamics, expected for both the molecular reorientations *and* the ionic hopping transport, can be separately detected. A spectral relaxation feature due to ion hopping is expected, based on the findings for materials without dipole-active reorientations[60,62,65] and theoretically predicted within certain ion-hopping models.[70,71] However, in most materials containing reorienting molecules, its unequivocal detection is hampered by the dominating dipolar relaxation process.[53] This is especially the case in liquid electrolytes like ionic liquids[53] or deep eutectic solvents,[55] where the (often neglected) reorientational and ionic motions should exhibit similar time scales because both are coupled to the viscosity. In contrast, in the present PC electrolytes, a separate analysis becomes possible due to the complete decoupling of dipolar and ionic motions.

## ACKNOWLEDGEMENTS

This work was supported by the Deutsche Forschungsgemeinschaft (grant No. LU 656/3-1).

The data that support the findings of this study are available from the corresponding author upon reasonable request.

9